\documentclass[10pt,letterpaper]{article}
\usepackage[top=0.85in,left=2.75in,footskip=0.75in]{geometry}
\usepackage{amsmath,amssymb}
\usepackage{changepage}
\usepackage[utf8x]{inputenc}
\usepackage{textcomp,marvosym}
\usepackage{cite}
\usepackage{nameref,hyperref}
\hypersetup{
	colorlinks=true,
	linkcolor=blue,
	filecolor=magenta,      
	urlcolor=cyan,
}
\usepackage[right]{lineno}
\usepackage{microtype}
\DisableLigatures[f]{encoding = *, family = * }
\usepackage[table]{xcolor}
\usepackage{array}
\usepackage{float}
\usepackage{mathtools}
\newcolumntype{+}{!{\vrule width 2pt}}

\newlength\savedwidth

\raggedright
\usepackage[aboveskip=1pt,labelfont=bf,labelsep=period,justification=raggedright,singlelinecheck=off]{caption}

\makeatletter
\renewcommand{\@biblabel}[1]{\quad#1.}
\makeatother

\usepackage{lastpage,fancyhdr,graphicx}
\usepackage{epstopdf}
\fancyheadoffset[L]{2.25in}
\fancyfootoffset[L]{2.25in}
\def\R{\mathbb{R}}

\DeclareMathOperator{\Ima}{Im}
\DeclareMathOperator{\Arg}{Arg}
\def\d{\mathrm{d}\,}
\newcommand{\beginsupplement}{
	\setcounter{table}{0}
	\setcounter{figure}{0}
	\setcounter{section}{0}
	\setcounter{equation}{0}
	\setcounter{paragraph}{0}
	\renewcommand{\thetable}{S\arabic{table}}
	\renewcommand{\thefigure}{S\arabic{figure}}
	\renewcommand{\theparagraph}{S\arabic{paragraph}~Text}
	\renewcommand{\theequation}{S\arabic{equation}}
	\renewcommand{\thesubsection}{S\arabic{subsection}}
	\renewcommand{\thesubsubsection}{\thesubsection.\arabic{subsubsection}}
	\setcounter{secnumdepth}{4}
}

\begin{document}
\vspace*{0.2in}

\begin{flushleft}
{\Large\textbf{Topological portraits of multiscale coordination dynamics} }
\newline
\\
Mengsen Zhang\textsuperscript{1,2*\dag},
William D. Kalies\textsuperscript{3},
J. A. Scott Kelso\textsuperscript{1,4},
Emmanuelle Tognoli\textsuperscript{1}
\\
\bigskip
\textbf{1} Center for Complex Systems and Brain Sciences, Florida Atlantic University, Boca Raton, Florida, USA
\\
\textbf{2} Department of Psychiatry and Behavioral Sciences, Stanford University, Stanford, California, USA
\\
\textbf{3} Department of Mathematical Sciences, Florida Atlantic University, Boca Raton, Florida, USA
\\
\textbf{4} Intelligent System Research Centre, Ulster University, Derry$ \sim $Londonderry, Northern Ireland
\\
\bigskip

* mengsenz@stanford.edu
\\
{\dag} this author is now with Stanford University

\end{flushleft}

\section*{Abstract}
Living systems exhibit complex yet organized behavior on multiple spatiotemporal scales. To investigate the nature of multiscale coordination in living systems, one needs a meaningful and systematic way to quantify the complex dynamics, a challenge in both theoretical and empirical realms. The present work shows how integrating approaches from computational algebraic topology and dynamical systems may help us meet this challenge. In particular, we focus on the application of multiscale topological analysis to coordinated rhythmic processes. First, theoretical arguments are introduced as to why certain topological features and their scale-dependency are highly relevant to understanding complex collective dynamics. Second, we propose a method to capture such dynamically relevant topological information using persistent homology, which allows us to effectively construct a multiscale topological portrait of rhythmic coordination. Finally, the method is put to test in detecting transitions in real data from an experiment of rhythmic coordination in ensembles of interacting humans. The recurrence plots of topological portraits highlight collective transitions in coordination patterns that were elusive to more traditional methods. This sensitivity to collective transitions would be lost if the behavioral dynamics of individuals were treated as separate degrees of freedom instead of constituents of the topology that they collectively forge. Such multiscale topological portraits highlight collective aspects of coordination patterns that are irreducible to properties of individual parts. The present work demonstrates how the analysis of multiscale coordination dynamics can benefit from topological methods, thereby paving the way for further systematic quantification of complex, high-dimensional dynamics in living systems.

\section*{Introduction}
A complex system (e.g. a biological, social, ecological system) is often bound together by the coordination between many dynamic processes at multiple spatiotemporal scales \cite{Waddington1962,Simon1977,Buzsaki2006,Lemke2000,Holling2001,Nathan2008,Aguilera2018} --- a multiscale coordinative structure \cite{Kelso2009}. Yet when faced with such multiscale dynamics, we find ourselves short of proper tools to describe them in a way that does justice to all relevant scales. In the present work, we propose a topological approach to analyzing dynamic patterns generated by multiscale coordinative structures. Topological methods have been shown to detect dynamic features of systems exhibiting, e.g., stable, spatiotemporal chaos \cite{Gameiro2004,Krishnan2007,Kramar2016}. Here we leverage existing computational topology tools, principally, persistent homology to capture the scale-dependency of topological features hidden in complex coordination patterns and detect transitions between them.  

The quantitative and systematic study of multiscale coordinative structures requires data analytic tools that are tuned to capture dynamic features across scales, that is, without predefining a specific scale of analysis or using a different system of measurement for different scales. The development of such tools is challenging since the measurement not only needs to be multiscale in nature, but also dynamically meaningful: it should capture dynamic patterns of coordination and the transition between coordination patterns in time \cite{Kelso1984,Kelso2009}. In the present paper, we propose a \textit{multiscale topological} approach to this problem. Instead of keeping track of individual state variables, we study the topological features of the spatiotemporal patterns generated by virtue of their interaction. Although our approach is customized for the study of phase coordination between multiple rhythmic processes, it also provides a prototype for general coordination problems. We explore the conditions under which multiscale dynamics emerge in rhythmic coordination, the difficulties that arise in analyzing such data, and why a multiscale topological approach helps to resolve them.

The study of rhythmic coordination has been an essential part of understanding collective dynamics in complex systems like the brain \cite{Kelso1995,Varela2001,Bressler2001,Buzsaki2006,Bressler2006,Kelso2009,Tognoli2009,Kelso2012,Tognoli2014} and groups of humans or other animals \cite{Winfree1967,Zerubavel1985,Schoner1990,Neda2000PRE,Lagarde2005,Alderisio2017,Zhang2018,Tognoli2018}. Two theoretical mechanisms are often used to explain empirically observed phase coordination phenomena, namely, phase-locked synchronization (e.g. \cite{Kuramoto1984,Winfree2001geometry}) and metastable coordination dynamics (e.g. \cite{Kelso1995,Tognoli2009,Kelso2012,Tognoli2014}). If all oscillators of a system are phase-locked so that phase relations are constant over time, they share a common instantaneous frequency. In this case, multiscale behavior is not possible. In metastable coordination, phase relations are not permanent but formed intermittently, i.e.~oscillators dwell at certain preferred phase relations temporarily when they pass by them (further illustrated in \nameref{section:methods}). In this case, oscillators do not converge to the same frequency over time so that the individuality or diversity of the component oscillators is somewhat preserved. It is this preservation of diversity during metastable coordination that gives rise to the coexistence of multiple spatiotemporal scales, e.g. phase relations form intermittently at different rates and among groups of different sizes. 

Metastable coordination can become quite obscure when the system involves many coupled oscillators, and frequency diversity is large. Moreover, traditional methods that work well in low-dimensional, low-diversity settings,
may be rendered less effective. For the rest of the Introduction, we demonstrate this point through two example trials of rhythmic social coordination from a human experiment \cite{Zhang2018} (Section \nameref{section:examples}). Metastable coordination of low dimensionality and diversity can be directly interpreted by visual examination of the relative phase dynamics (Fig~\ref{fig:triad_example}A), but this is far more difficult when the dimensionality and diversity are high (Fig~\ref{fig:eight_example}A). Further, we show that recurrence plots, a classical phase space method for nonlinear dynamical systems \cite{Eckmann1987,Marwan2007,Kantz2003}, do not improve our understanding of the high-dimensional, high-diversity example (Section \nameref{section:recur_phi} and Fig~\ref{fig:recur_phi}).

\subsection*{Metastable dynamics of human social coordination}\label{section:examples}
\begin{figure}[H]
	\begin{adjustwidth}{-2.25in}{0in}
	\centering
	\includegraphics[width=\linewidth]{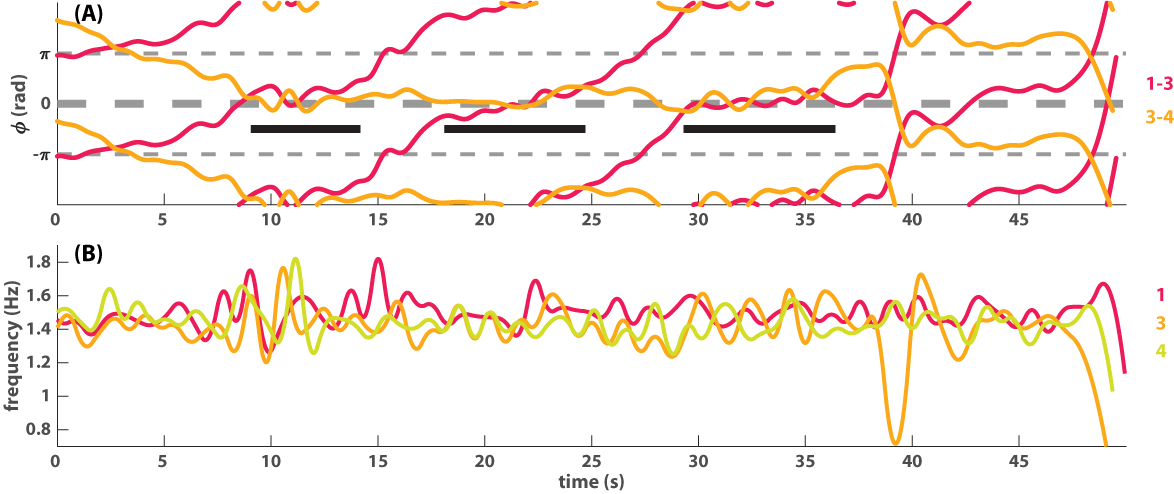}
	\caption[An example of triadic coordination dynamics.]{{\bf An example of triadic coordination dynamics}. Coordination among three agents (labeled as 1, 3, and 4) is shown as the dynamics of two pairwise relative phases (A) and three instantaneous frequency trajectories (B). Around 10s, three agents formed an all-inphase relation ($ \phi_{1,3}\approx\phi_{3,4}\approx0 $ rad) for a few seconds, marked by a black bar on the left in (A). This pattern recurred intermittently two more times (middle, right bars in A), which ended when pair 3-4 switched to antiphase (40-48s, orange trajectory $ \phi_{3,4}\approx \pi $ rad). Both relative phase trajectories (A) evolve on a slow time scale because the frequencies of these three agents are very close (B).}\label{fig:triad_example}
	\end{adjustwidth}
\end{figure}

\begin{figure}[H]
	\begin{adjustwidth}{-2.25in}{0in}
	\centering
	\includegraphics[width=\linewidth]{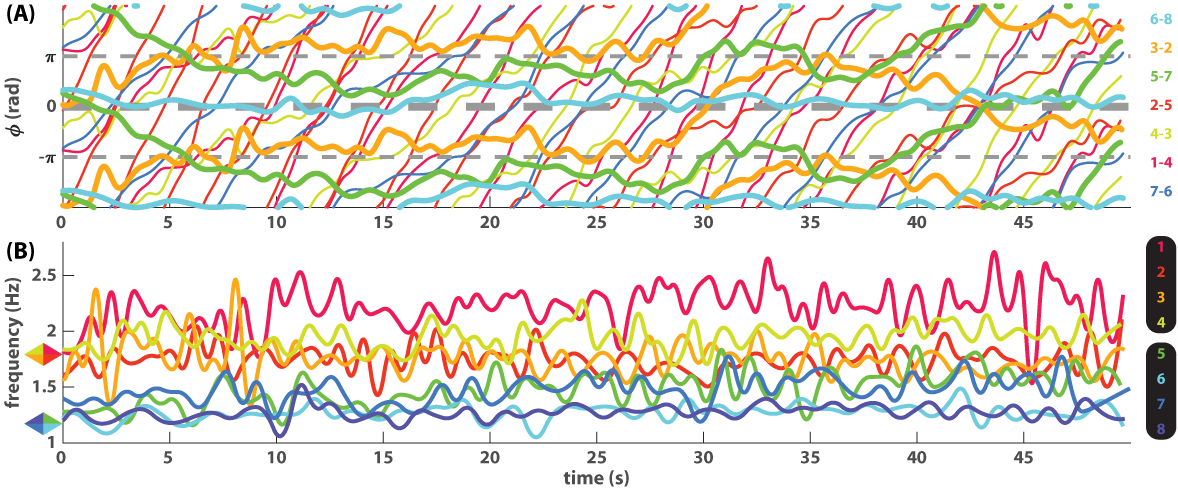}
	\caption[An example of eight-agent coordination dynamics.]{{\bf An example of eight-agent coordination dynamics shown as seven pairwise relative phases (A) and eight instantaneous frequency trajectories (B)}. In (A), slowly varying phase relations are shown as thick lines (orange trajectory 3-2, green 5-7, cyan 6-8), whereas fast varying phase relations are shown as thin lines with much steeper slopes than the thick lines. In (B), the corresponding frequency trajectories indicate that frequency diversity is much greater than in Fig~\ref{fig:triad_example}B. Because agents were paced with two different metronomes prior to interaction (1.2 and 1.8 Hz, marked by colored rhombi on the left), the ensemble of eight starts off as two frequency groups, one in warm colors (1, 2, 3, 4) and one in cold colors (5, 6, 7, 8). Toward the end of the trial, members from the two groups begin to mingle.}\label{fig:eight_example}
	\end{adjustwidth}
\end{figure}

Fig~\ref{fig:triad_example} and \ref{fig:eight_example} provide two examples of actual metastable coordination between three and eight persons respectively. They show the general form of metastable phase relations and, by virtue of the contrast between the two, reveal why high-dimensional metastable coordination is difficult to analyze. Later on, we use them to validate and test the topological approach proposed in the following sections. 

The two examples shown in Fig~\ref{fig:triad_example} and \ref{fig:eight_example} were recorded in two separate trials in an experiment of human social coordination, dubbed the ``Human firefly" experiment \cite{Zhang2018}. In the experiment, multiple human subjects tapped together rhythmically on a set of touch pads. Taps by one person can be seen by others in real time as flashes of specific LEDs (see \cite{Zhang2018} for details), creating the possibility of subjects coordinating with each other spontaneously (i.e. they were not explicitly instructed to do so). Each subject is here referred to by a number from one to eight, associated with a specific touch pad and LED assignment. Diversity in the tapping frequencies was manipulated by pacing each subject for 10s with a separate metronome before they saw each other's behavior for 50s. The pacing frequency was always the same within the group of subjects numbered 1 to 4, and within the group of subjects numbered 5 to 8, but could be different between the two groups. 

In the first example (Fig~\ref{fig:triad_example}), we show the coordination dynamics between three agents paced with metronomes of the same frequency, which may be interpreted visually through two pairwise relative phases, a crucial coordination variable \cite{Haken1985,Kelso2012}. In Fig~\ref{fig:triad_example}A, coordination among three agents, numbered 1, 3, and 4, is shown in terms of two phase relations (1-3 magenta and 3-4 orange). From 10s to 40s, the system dwells recurrently at an all-inphase pattern, where all taps are aligned in time (duration marked by three black bars; between the bars only 3-4 are inphase with agent 1 wrapping). Then the behavior switches to a partly inphase, partly antiphase pattern after 40s (1-3 inphase and 3-4 antiphase) as in Fig~\ref{fig:triad_example}A. This is due to a sudden slowing down at around 40s of agent 3, shown as the orange trajectory in Fig~\ref{fig:triad_example}B (Note that the frequency is the time derivative the phase divided by $ 2\pi $). The relative ease of interpretation comes from the facts that the number of interacting agents is small (low-dimensionality) and that they stay close in frequency (Fig~\ref{fig:triad_example}B). As a result, their phase coordination occurs on visually comparable time scales. 

This ease of analysis is lost when more interacting agents and frequency diversity are involved, as illustrated in the second example (Fig~\ref{fig:eight_example}). Here eight agents were paced with different metronomes before interaction, four at 1.2 Hz, four at 1.8 Hz (see caption of Fig~\ref{fig:eight_example}). The dynamics of pairwise relative phases (Fig~\ref{fig:eight_example}A) is much more difficult to decipher now that behavior is evolving on very different time scales. Observe the slow dynamics shown as thickened trajectories, mostly horizontal (e.g. cyan trajectory for pair 6-8), in contrast to fast dynamics shown as thin trajectories, mostly wrapping, i.e.~with a steep slope (e.g. blue trajectory for pair 7-6). Even though each trajectory can be singled out and studied carefully in separation, it remains unclear how these multiple phase relations constrain each other and perhaps form higher-level structures. On the other hand, the frequency dynamics (Fig~\ref{fig:eight_example}B) is more informative regarding the global organization: eight agents are separated into two frequency groups at the beginning, coded in warm vs.~cold colors, but gradually become intermingled over the course of the trial. Yet it is not apparent how to relate such a global trend in frequency to underlying phase coordination. Clearly, to characterize this kind of multiscale coordination dynamics requires additional computational tools. 

\subsection*{Limitation of traditional recurrence plots in explicating multiscale coordination dynamics}\label{section:recur_phi}
\begin{figure}[H]
	\begin{adjustwidth}{-2.25in}{0in}
		\centering
		\includegraphics[width=\linewidth]{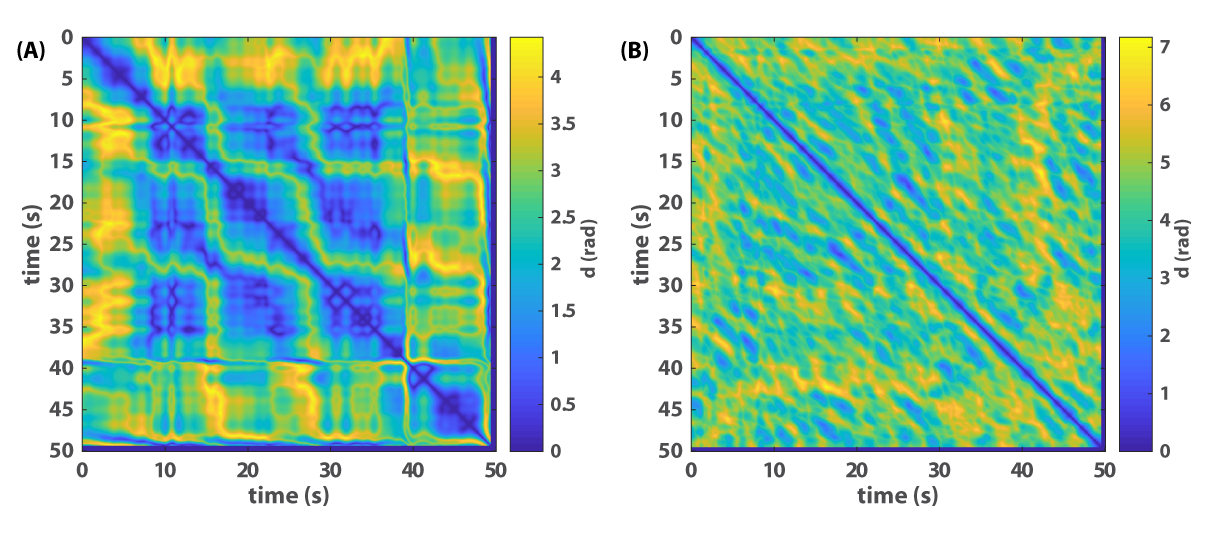}
		\caption[Recurrence plots of relative phase dynamics.]{{\bf Recurrence plots of relative phase dynamics}. (A) shows the recurrence plot of phase relations among three agents where the state variable is a vector with the 2 relative phases shown in Fig~\ref{fig:triad_example}A as components, and (B) recurrence plot of eight agents where the state variable is a vector of the 7 relative phases plotted in Fig~\ref{fig:eight_example}A as components. }\label{fig:recur_phi}
	\end{adjustwidth}
\end{figure}

A recurrence plot \cite{Eckmann1987,Marwan2007} is a powerful tool for visualizing and analyzing patterns of nonlinear dynamical systems, especially when the state space itself is too high-dimensional to visualize. Rather than showing the state variable itself, it shows the relation between states at different points in time, e.g. as a distance matrix, from which one can infer how frequently a system visits different points in the state space. In Fig~\ref{fig:recur_phi}A-B, we show the recurrence plots of the two examples above in terms of the state variable $ \vec{\phi}(t) $, whose components are relative phases shown in Fig~\ref{fig:triad_example}A and \ref{fig:eight_example}A respectively. The components of the distance matrices are defined as 
\begin{equation}
d_{t_1,t_2}= \| W\big(\vec{\phi}(t_1)-\vec{\phi}(t_2)\big) \| \label{eq:distphi}
\end{equation}
where function $ W $ wraps each component to the interval $ (-\pi,\pi] $ by $W(z)_k=\Arg(e^{iz_k})$ and $ \|\cdot\| $ is the $ L_2 $-norm.

The recurrence of relative phases (Fig~\ref{fig:recur_phi}A) clearly captures the dynamic structure of the three-agent example (Fig~\ref{fig:triad_example}A). In Fig~\ref{fig:recur_phi}A, a 3-by-3 grid between 10 and 40s reflects the main recurrent pattern, and the blocks before 10s and after 40s reflect two other patterns. In contrast, as shown in Fig~\ref{fig:recur_phi}B, not much useful information is revealed in the eight-agent case. 

Fig~\ref{fig:recur_phi} illustrates that conventional recurrence plots can be useful for low-dimensional coordination dynamics but may not work so well for high-dimensional dynamics involving multiple spatiotemporal scales. In order to shed more light on the latter, we next propose a topological way of studying metastable patterns. Subsequently, we present a computational method for constructing a topological recurrence plot, inspired by \cite{Kramar2016}, which captures the dynamics of topological features, such as connected components and loops (holes), in the collective patterns that are irreducible to the properties of the individual components. 

\section*{Multiscale topological portraits and topological recurrence}\label{section:methods}
Here we propose a new approach to the study of metastable coordination patterns by means of their topological features. More specifically, we construct multiscale topological portraits of coordination patterns and study the recurrence plot of said portraits, i.e. a topological recurrence plot. Further, we show how prominent transitions observed in a topological recurrence plot reveal the time of transitions in actual coordination patterns in terms of relative phases and instantaneous frequencies. 

In the following sections, we first explain intuitively what we mean by multiscale topological portraits  (Section \nameref{section:method_topportrait}), and then theoretically justify their relevance to metastable coordination patterns (Section \nameref{section:method_topmeta}). Finally, we give a technical description of the construction of multiscale topological portraits (Sections \nameref{section:method_decomp} and \nameref{section:method_tda}) with associated recurrence plots (Section \nameref{section:method_toprecur}). 

\subsection*{Multiscale structures and their topological portraits} \label{section:method_topportrait}
\begin{figure}[H]
	\begin{adjustwidth}{-2.25in}{0in}
		\centering
		\includegraphics[width=\linewidth]{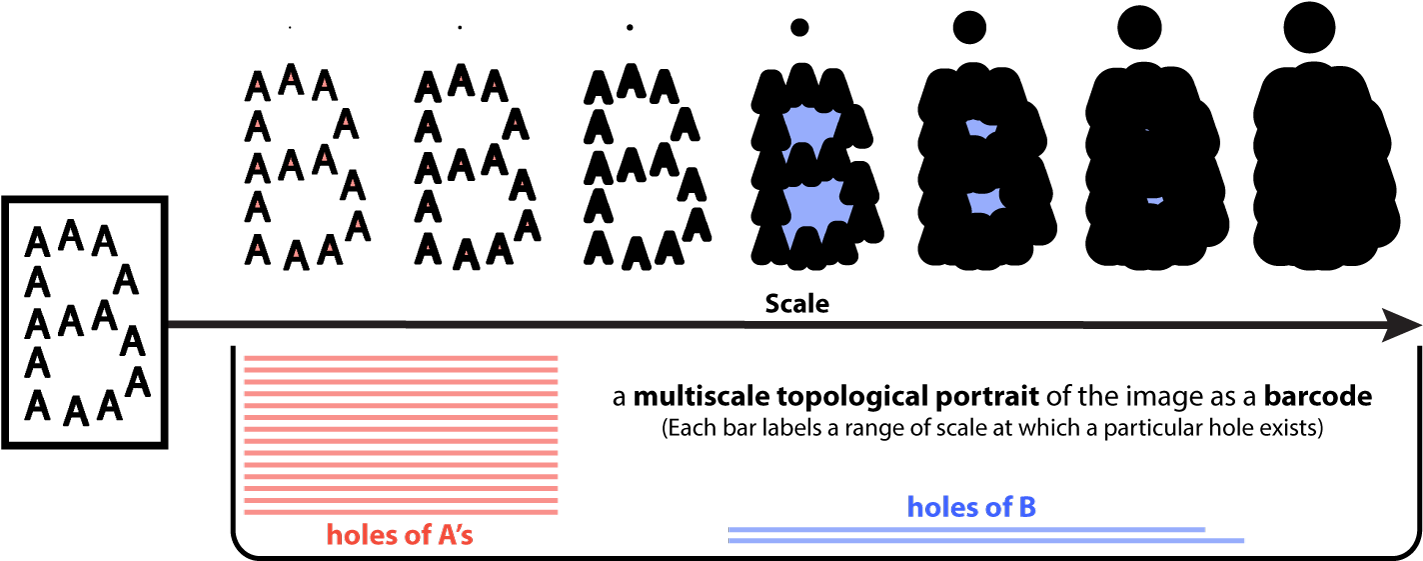}
		\caption{
			{\bf A big letter B made up of many small letter A's and its multiscale topological portrait}. This figure illustrates intuitively how one may gain insights about a multiscale structure, i.e.~a B of many A's (image in left box), by studying the scale-dependency of its topological features, i.e. connected components and holes. The image may be viewed at succeeding greater scales as a sequence of images (above the scale axis) created by blowing up each pixel to a disk with increasing radius (size of disks are shown on top). At finer scales (left two images above the scale axis), the image has 14 connected components and 14 holes (from 14 A's). At intermediate scales (4\textsuperscript{th}--6\textsuperscript{th} image from left above the scale axis), a large connected component is formed with two holes in it (from one B). At even larger scales, all holes are filled in (right-most image above the scale axis). A multiscale topological portrait (below the scale axis) summarizes the emergence and disappearance of holes as a function of scale. The portrait consists of a collection of bars (a barcode), where each bar represents a particular hole and the scales at which it exists. The portrait captures the separation between two descriptions (many A's or a B) in scale, by capturing the scale-dependency and relative size of topological features (14 red bars appear only at finer scales and are shorter, reflecting A's; blue bars only appear at grosser scales and are longer, reflecting the B; one blue bar is longer than the other because the lower loop in B is larger than the upper). This portrait is drawn by hand for illustrative purposes, a computed version based on persistent homology (see text) is shown in Fig~\ref{SIfig:ab_pd}~B.
		}\label{fig:ab}
	\end{adjustwidth}
\end{figure}

\begin{figure}[H]
	\begin{adjustwidth}{-2.25in}{0in}
		{\centering
			\includegraphics[width=\linewidth]{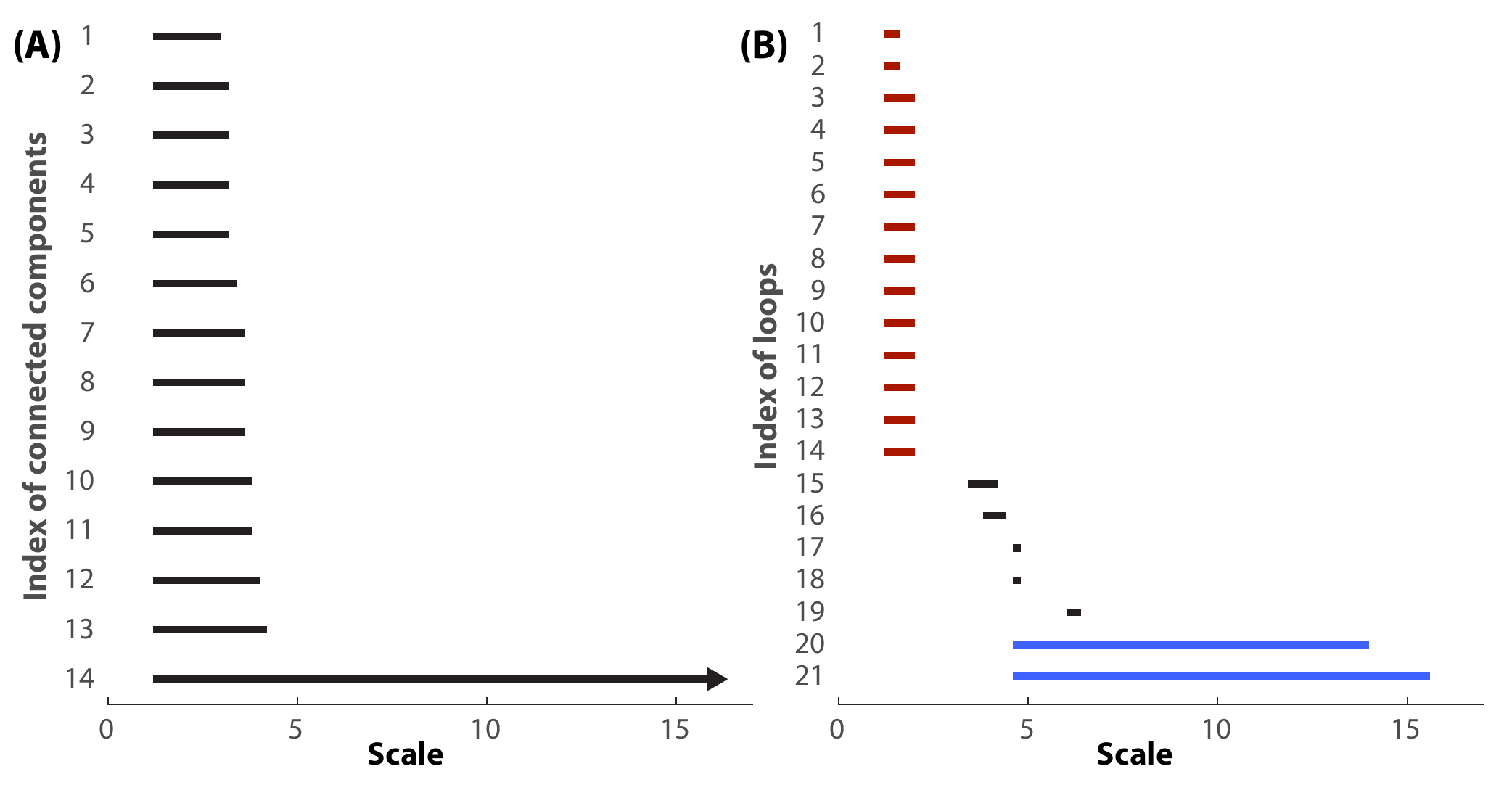}}
		\caption{{\bf Persistent homology of a big letter B of many small letter A's shown as barcodes.} The 0\textsuperscript{th} and 1\textsuperscript{st} persistent homology characterize the scale-dependency of connected components (A) and loops (B) respectively, computed from an image of a big letter B of many small letter A's (Fig~\ref{fig:ab} left box; the image used for computation has $ 61\times 100$ pixels). In (A), each bar represents a particular connected component, whose left end indicates at which scale this connected component emerges, and the right end at which scale it disappears (by merging into a larger component). There are clearly 14 connected components at finer scales (scale, i.e. radius of the disk centered at each pixel, below 5) and 1 at grosser scales (right arrow in bar 14 indicates that this component never disappears with increasing scale). In (B), each bar represents a loop. There are clearly 14 loops at finer scales (below 3; red bars) and two major loops at grosser scales (blue bars; some more transient features also appear in the transition between the 14-loop and the 2-loop configuration, which are also features emerged during scaling of the image; colors are only for highlights, not computed). Together, these two multiscale topological portraits (A-B) faithfully capture the coexistence of two descriptions of the image (many A's or one B) and their separation in scale.}\label{SIfig:ab_pd}
	\end{adjustwidth}
\end{figure}

By a multiscale structure, we refer to a spatial or spatiotemporal structure the description of which is scale-dependent. For example, a letter B made up of many A's as in Fig~\ref{fig:ab} (image in left box) may be described as a collection of A's at finer scales or a single B at grosser scales. Both descriptions are correct at their respective scales, and together they form a more complete portrayal of the structure, i.e.~a multiscale portrait. Such a portrait should capture the scales at which each description is characteristic. One way to keep track of these descriptions across scales is through topological features--- an A has one connected component and one loop (hole), and a B has one connected component and two loops. Thus, in a B consisting of many A's, one should find 14 connected components and 14 loops at finer scales, but only one connected component and two loops at grosser scales. To create scale-dependent versions of the image (Fig~\ref{fig:ab} left box), we can replace each pixel of the image with a disk of growing radius (upper panel of Fig~\ref{fig:ab}, smaller scales/radius on the left, larger on the right). Graphically, we can use a bar to mark the scales at which each topological feature exists, e.g.~red bars in Fig~\ref{fig:ab} representing loops from A's. The collection of all such bars, a barcode, can serve as a \textit{multiscale topological portrait} of the structure. Fig~\ref{fig:ab} lower panel shows such a portrait, which indicates that the many-A description only exists at finer scales (many red bars on the left), and the one-B description only exists at grosser scales (two blue bars on the right). In practice, \textit{persistent homology} \cite{Zomorodian2005, Carlsson2009,Ghrist2007} is a natural choice for computing such a multiscale topological portrait. This homology not only captures certain topological features of the structure at each scale, but it also describes how these features persist across scales. Fig~\ref{SIfig:ab_pd}~B shows the actual barcode of loops computed using persistent homology, which corresponds well to the intuition conveyed in Fig~\ref{fig:ab} lower panel; in addition, Fig~\ref{SIfig:ab_pd}~A shows the barcode of connected components, which captures the existence of 14 components at finer scales, and 1 at grosser scales. Before giving a technical description of persistent homology (Section \nameref{section:method_tda}), we first address whether metastable patterns are multiscale structures that can be suitably characterized, in principle, by multiscale topological portraits. 

\subsection*{Multiscale topological features of metastable patterns}\label{section:method_topmeta}

\begin{figure}[H]
		\centering
		\includegraphics[width=\linewidth]{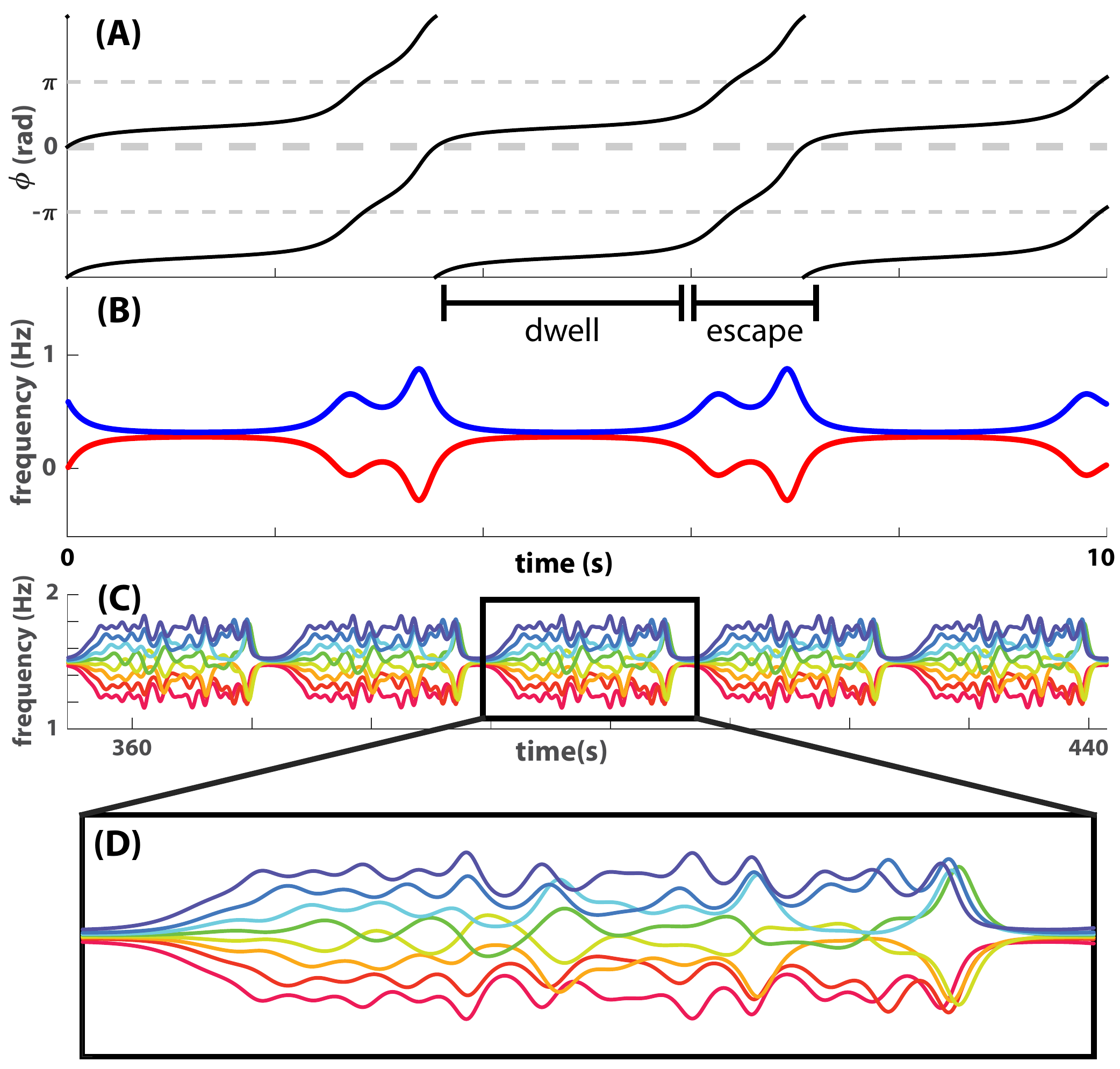}
		\caption{
			{\bf Metastable patterns characterized by topological features in frequency graphs}. Examples of simulated coordination dynamics are presented here to show why multiscale topological features are relevant to metastable patterns. (A-B) give a simple illustration of how relative phase dynamics (A) can be studied in terms of topological features in its corresponding frequency graph (B) defined as the collection of instantaneous frequency trajectories. A dwell in the relative phase (A; period labeled as ``dwell") is reflected as the merging of corresponding frequency trajectories into a single connected component (B, observed at a sufficiently gross scale). An escape in the relative phase (A; period labeled as ``escape") is reflected as the branching of frequency trajectories into two connected components or the formation of a loop if viewed in an extended time window (e.g. a window centered around the escape that extends to the middle of neighboring dwells). Thus, the dynamics of metastable phase coordination can be studied as topological features in the corresponding 2-dimensional frequency graph, which is very convenient when the dimension of the dynamical system increases. (C) shows the frequency graph of metastable coordination between eight oscillators, whereas a zoomed-in version of one period of the pattern is shown in (D). For such a complex pattern (D), the spacing between curves and the size of loops are very diverse, reflecting dwell-escape dynamics at various spatiotemporal scales. As a result, topological features in the frequency graph have to be measured at multiple scales to capture the complexity of such metastable patterns.
		}\label{fig:explain_freqtopo}
\end{figure}

How can we relate the multiscale topological features illustrated in the above example  to dynamic metastable patterns? The answer lies in the connection between dynamics of relative phases and that of instantaneous frequencies, given by the time derivative of phase. Due to this mathematical relation between phase and frequency, the dwell-escape dynamics central to metastable phase coordination corresponds to specific topological features in the frequency graph, here defined as the collection of all instantaneous frequency trajectories. This point is illustrated in Fig~\ref{fig:explain_freqtopo} using numerically simulated metastable patterns of a model \cite{Zhang2019jrsi} derived specifically from the ``Human firefly" experiment \cite{Zhang2018} (see also \cite{Zhang2018Dis} and \ref{SItext:simulation}). Fig~\ref{fig:explain_freqtopo}A shows a canonical example of dyadic metastable coordination in terms of the temporal evolution of relative phase. Dwells and escapes in the relative phase dynamics (Fig~\ref{fig:explain_freqtopo}A) correspond to different numbers of connected components and loops in the corresponding frequency graph (Fig~\ref{fig:explain_freqtopo}~B): an escape can be seen as a split of one connected component into two when viewed in a short time window or as the formation of a loop when viewed over an extended time window. Indeed, the distance between two frequency trajectories reflects the slope of the corresponding relative phase trajectory, which is by definition smaller during a dwell and greater during an escape. The size of a loop, measured as the area enclosed by two frequency trajectories during a period of escape, as in Fig~\ref{fig:explain_freqtopo}~B, reflects the amount of change in relative phase between two dwells. In the example, the relative phase changes by 1 cycle between two consecutive dwells at inphase; thus, the area of the corresponding loop in Fig~\ref{fig:explain_freqtopo}~B is a quantal value of one. 

It is important to notice that here the number of connected components or loops is in fact scale-dependent. At too fine a scale, the two curves in Fig~\ref{fig:explain_freqtopo}~B never cross so that there are two connected components and no loop. At too gross a scale, all loops are filled in so that there is one connected component and no loops. At intermediate scales, topological features can capture the dwell$\sim$escape dynamics characteristic of metastable coordination. When more oscillators coordinate together metastably, multiple characteristic scales may coexist due to the diversity of distance between trajectories and the size of loops they enclose, e.g.~as in Fig~\ref{fig:explain_freqtopo}~C-D; see \cite{Zhang2018Dis} for more examples. Multiscale topological portraits of frequency graphs are suitable for characterizing such complex metastable patterns in that they capture topological features that are relevant to the coordination dynamics across multiple scales. In the next few sections, we give more technical descriptions of how to construct multiscale topological portraits and topological recurrence plots from real data. 

\subsection*{Preparing empirical data for topological analysis}\label{section:method_decomp}
In the previous section, we motivate the use of topological analyses on frequency graphs as a way to characterize metastable coordination dynamics. In the ideal case of simulated data, one may directly use the frequency graph, or segments of it, to construct multiscale topological portraits. If the parameters of the dynamical system are stationary (e.g. $\omega_i$, $a$, $b$ in equation~\ref{eqn:NHKB}), the same metastable pattern is repeated over time as in Fig~\ref{fig:explain_freqtopo}C. Such ideal scenarios cannot be assumed for empirical data such as the examples of human social coordination dynamics illustrated above. In empirical data, transitions between different metastable patterns are possible. During such a transition, instantaneous frequency trajectories may oscillate significantly forming transient connected components or loops that do not reflect the dwell-escape dynamics of a stationary metastable pattern (see e.g.~oscillation of orange trajectory 3 in Fig~\ref{fig:triad_example}~B around 40s). For this reason, we first decompose experimental data into slow and very fast time scales, assuming that metastable patterns unfold at the slow time scale, which is a basic assumption in the theoretical study of phase coordination, cf.~\cite{Winfree1967,Kuramoto1984,Haken1985}. 

\begin{figure}[H]
	\begin{adjustwidth}{-2.25in}{0in}
		\centering
		\includegraphics[width=\linewidth]{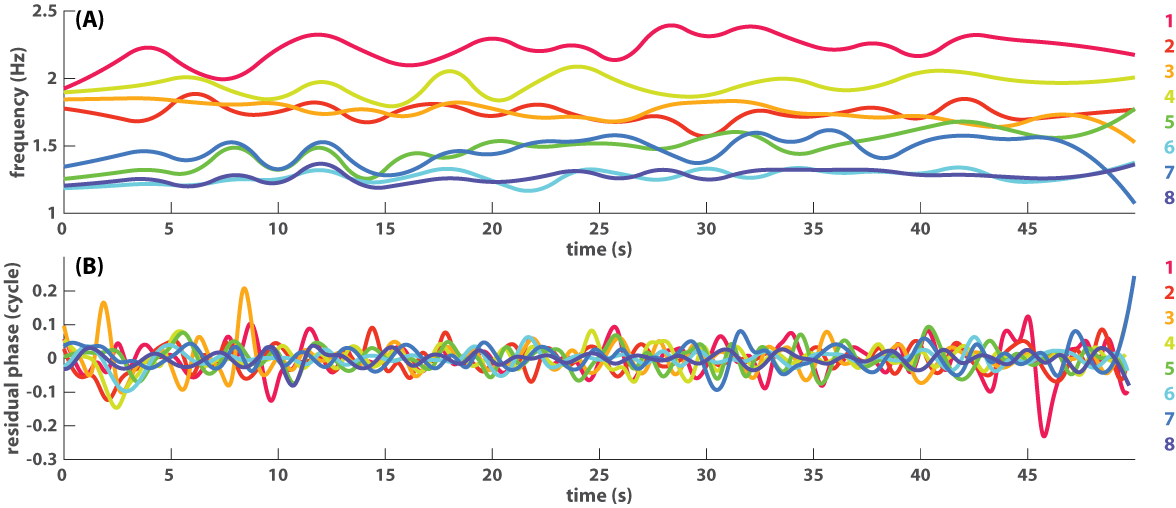}
		\caption[Decomposition of absolute phase dynamics.]{{\bf Decomposition of absolute phase dynamics for the eight-agent example (Fig~\ref{fig:eight_example})}. The absolute phase of each agent is decomposed into to a slowly varying frequency component (A) and a fast varying residual phase (B).}\label{fig:eight_decomp}
	\end{adjustwidth}
\end{figure}

Specifically, the phase  $ \varphi $ of each agent was decomposed into a slowly varying frequency component $ \hat{f} $ in Hz (e.g. Fig~\ref{fig:eight_decomp}A for the eight-agent example) and a fast varying residual phase $ \varphi_r $ in cycles (Fig~\ref{fig:eight_decomp}B), without losing any information, i.e.~the original phase can be recovered by $ \varphi=2\pi\,\big( \int_{0}^{t}\hat{f}(\tau)\d \tau+ \varphi_r(t) \big)$. To obtain this decomposition, we first performed a least-squares fit of the phase to a piecewise cubic spline $ \hat{\varphi} $ using \texttt{splinefit} in Matlab with robust fitting parameter $ \beta=0.5 $ \cite{splinefit}. Knots of the spline were chosen at 2s intervals, based on the observation that dyadic phase coordination mostly exceeded 2s (i.e. 87\% of the dwells observed in the human experiment exceeded 2s. See the distribution of dwell times in Fig B of S1 File in \cite{Zhang2018}). The slow component (frequency) is the time derivative of $ \hat{\varphi}$, i.e. $ \hat{f}:=\frac{1}{2\pi}\frac{d\hat{\varphi}}{dt} $, and the fast component (residual phase) is $ \varphi_r:=(\varphi-\hat{\varphi})/2\pi $. Based on this decomposition, we can subsequently compute the multiscale topological portraits of 3-dimensional frequency-phase graphs with the added dimension of residual phase, instead of 2-dimensional frequency graphs. Note that parameters of the decomposition should, in general, be chosen based on properties of each specific dataset and tested on part of the dataset where the transitions between dynamic patterns are transparent (e.g. the triadic example in our case). 

To study the dynamics, we segment each frequency-phase graph into 2s windows (consistent with the decomposition above) such that consecutive windows overlap by 1s. In each window, each agent's behavior is then sampled at specific times to obtain a point cloud in 3-dimensional space, which is a set of $ M $ points whose coordinates correspond to local time, residual phase, and frequency respectively (e.g. in Fig~\ref{fig:pd}A, each point in the point cloud is shown as a small ball). For this study, $ M = 160 $ as each of the eight agents are sampled at 20 time points (at 0.1s intervals). Specifically, $ X(t)=\{x_1(t),\cdots, x_i(t),\cdots,x_M(t)\} $ where $x_i(t) \,\in\, U_t=I_t\times S^1 \times \R^+ $ with time interval $ I_t=[t-w/2,t+w/2] $ centered at time $ t $ of length $ w $, $S^1$ the set of possible residual phases, and $\R^+$ the set of possible frequencies. To later compute the topological portraits of each point cloud $ X(t) $, we equip the space $ U_t $ with the following metric. The distance between points $ a=(a_1,a_2,a_3)^\intercal$ and $b=(b_1,b_2,b_3)^\intercal \in U_t$ is given by
\begin{equation}
d(a,b)=\left\|\left(a_1-b_1,\frac{W(2\pi(a_2-b_2))}{2\pi}, a_3-b_3\right)^\intercal\right\| \label{eq:3dmetric}
\end{equation} 
where the function $ W $ wraps its argument to the interval $ (-\pi,\pi] $ by $W(z)=\Arg(e^{iz})$ and $ \|\cdot\| $ is the $ L_2 $-norm.

\subsection*{Persistent homology}\label{section:method_tda}
Persistent homology is a tool from  algebraic topology that captures ``holes" of a space at multiple scales. It was initially developed to understand the relative importance of topological features in data \cite{zomorodian2001,Edelsbrunner2002} rather than to characterize multiscale dynamic patterns. Yet as alluded to previously, persistent homology happens to be a natural tool for extracting dynamically-relevant information in complex rhythmic coordination patterns. Here we use it to construct a multiscale topological portrait of such patterns. In this section, we give a brief description of the persistent homology of frequency-phase graphs. For a formal account of homology, see \cite{Hatcher2001}, and persistent homology, see \cite{Zomorodian2005,Carlsson2009,Mischaikow2013}. 

Roughly speaking, the homology of a space counts ``holes" of different dimensions by aggregating local connectivity information into global invariants. The dimension of a hole is determined by the dimension of the boundary that encloses it, e.g. a 0-dimensional hole is a connected component, a 1-dimensional hole is a loop (as in an ``A"), and a 2-dimensional hole is a cavity (as in a basketball). This counting process is done algebraically, and computationally, which requires a finite, combinatorial description of the underlying space. 

\begin{figure}[H]
	\begin{adjustwidth}{-2.25in}{0in}
	\centering
	\includegraphics[width=\linewidth]{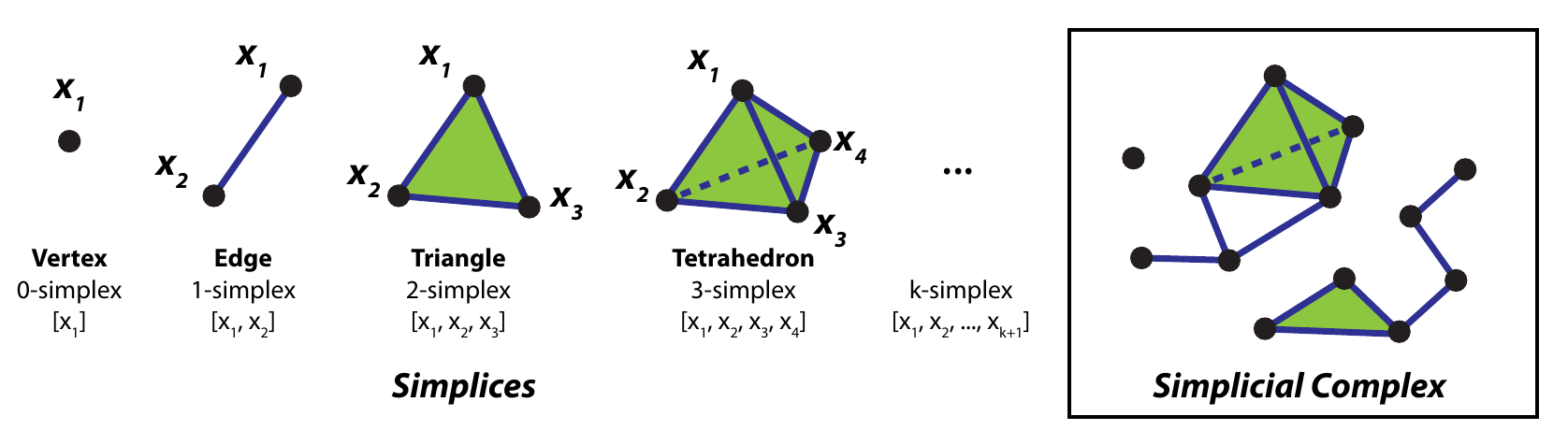}
	\caption[Simplices and a simplicial complex.]{{\bf Simplices and a simplicial complex}. Simplices are elementary geometric objects of different dimensionality, which can be combined into more complex structures, i.e. simplicial complexes. A $ k $-simplex can be thought of as a $ k $-dimensional triangle, determined by its $ (k+1) $ vertices. For example, a 2-simplex is a conventional triangle, determined by three vertices $ [x_1,x_2,x_3] $; a 0-simplex a vertex determined by itself $ [x_1] $; and a 1-simplex an edge determined by two vertices $ [x_1,x_2] $. A simplicial complex therefore can be described combinatorially as a set of vertices plus a collection of its subsets which represent higher-dimensional simplices connecting those vertices. }\label{fig:complex}
	\end{adjustwidth}
\end{figure}

\begin{figure}[H]
	\begin{adjustwidth}{-2.25in}{0in}
	\centering
	\includegraphics[width=\linewidth]{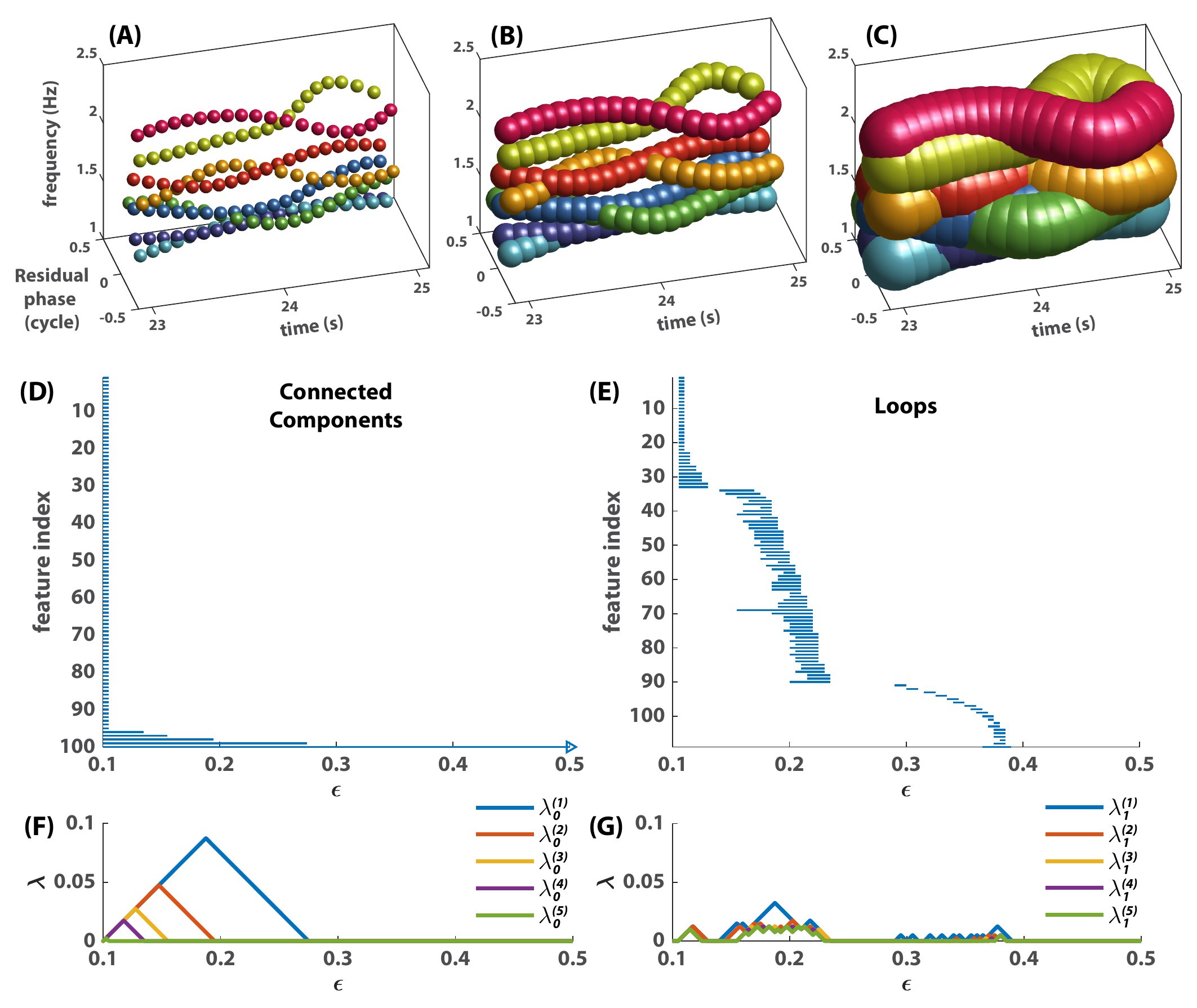}
	\caption[Persistence of topological features.]{{\bf Persistence of topological features}. (A-C) shows a coordination pattern represented by point clouds at three different scales $ \epsilon $, i.e.~a union of balls centered at each point with diameter $ \epsilon=0.1 $ in (A), $ \epsilon=0.2 $ in (B), or $ \epsilon=0.5 $ in (C). Note the merging of connected components from (A) to (B), the emergence of loops in (B) and their destruction in (C). The $ 0^{th} $ and $ 1^{st} $ persistent homology (connected components and loops respectively) of this point cloud are shown in (D,E) as barcodes, and in (F,G) as persistence landscapes. In (D,E), each horizontal bar represents a connected component (D) or loop (E), whose left (right) end indicates its birth (death) scale. Right arrow in (D) indicates that this component never dies (one connected component remains at any scale). (F,G) summarize the same information as a sequence of landscape functions, $ \lambda $, reflecting the most to least prominent homological features across scales (blue to green lines are the five largest landscape functions).  }\label{fig:pd}
	\end{adjustwidth}
\end{figure}

In the study of metastable patterns, the geometric structure of interest is a segment of a frequency-phase graph, which is a 3-D point cloud $ X =\{x_1,\cdots,x_M\}$ as described above. When measured at a specific scale $ \epsilon $, we obtain a union of balls centered at each point in $ X $, see Fig~\ref{fig:pd}A-C, i.e.~$X_\epsilon=\bigcup_{i=1}^M B_{\epsilon/2} (x_i) $. To algebraize the problem of finding homological features, we first map $ X_\epsilon $ to a \textit{simplicial complex}. The building blocks of a simplicial complex are \textit{simplices}, which can be thought of as triangles generalized to arbitrary dimensions (Fig~\ref{fig:complex}), i.e.~a $ k $-simplex is a $ k $-dimensional triangle spanned by its $ (k+1) $ vertices. A familiar example is a network, which is a simplicial complex containing only 0-simplices as vertices and 1-simplices as edges. In the present study, we construct the \textit{Rips complex} $ R_\epsilon(X) $ \cite{Hausmann1995} for each pattern $ X $ at scale $ \epsilon $, which is a computationally efficient proxy for $ X_\epsilon $. See \cite{Ghrist2007} for comparisons with other constructions. $ R_\epsilon(X) $ is an abstract simplicial complex consisting of all points in $ X $ as its vertices  and each $ k $-simplex whose vertices have all pairwise distances less than $ \epsilon $ where distance is defined in Eq~\ref{eq:3dmetric}.

 Algebraically, a simplicial complex is associated with a sequence of abelian groups $ C_k $, which are generated by the $ k $-simplices of $ R_\epsilon(X) $, along with corresponding linear operators $\partial_k\colon C_k\to C_{k-1},$ yielding
\begin{equation}
\cdots C_{k+1} \xrightarrow{\partial_{k+1}} C_k \xrightarrow{\partial_k} C_{k-1} \rightarrow \cdots  \rightarrow C_2 \xrightarrow{\partial_2} C_1 \xrightarrow{\partial_1} C_0 \xrightarrow{\partial_0} 0. \label{eq:chaincomplex}
\end{equation}
An element of $C_k$ is a finite, linear combination of $k$-simplices called a {\em $k$-chain.}
Each \textit{boundary operator} $ \partial_{k} $ maps a $ k $-simplex to a $(k-1)$-chain in its geometric boundary and
is defined so that $\partial_{k}\circ\partial_{k+1}=0$. 

The kernel of $\partial_k$ is the subgroup of $k$-chains whose boundary is trivial; this is the group of {\em $k$-cycles}, $\ker \partial_k$.
The image of $\partial_{k+1}$ is the subgroup of $k$-chains generated by the boundaries of $(k+1)$-simplices; this is the subgroup of {\em $(k+1)$-boundaries}, $\Ima\partial_{k+1}.$
Since $\partial_{k}\circ\partial_{k+1}=0,$ we have $\Ima\partial_{k+1}\subset\ker\partial_{k}$, and the \textit{$k$-th homology group} of the simplicial complex is defined by the quotient group
\begin{equation}
H_k=\ker \partial_k/\Ima \partial_{k+1} \label{eq:H_k}.
\end{equation}
$H_k$ algebraically describes the boundary-less $k$-chains that do not bound a $(k+1)$-chain, i.e.~$k$-dimensional ``holes", which can be counted by computing the number of independent free generators, or Betti number, of $H_k$. This heuristic (and simplistic) description of homology is a reasonable description in low-dimensions.

Persistent homology keeps track of each independent $ k $-dimensional hole in the  family of Rips complexes $ R_{\epsilon_i}$ in Eq~\ref{eq:filteredrips}, which may emerge at any scale, throughout its life span across scales $\epsilon_i$, where $0\leq\epsilon_{i}<\epsilon_{i+1} $ for any index $ 0\leq i<P-1 $.  
\begin{equation}\label{eq:filteredrips}
R_{\epsilon_0}(X)  \subset  R_{\epsilon_1}(X)  \subset\cdots\subset  R_{\epsilon_{i-1}}(X)  \subset  R_{\epsilon_i}(X)  \subset  R_{\epsilon_{i+1}}(X) \subset \cdots \subset  R_{\epsilon_P}(X)
\end{equation}
\begin{equation}\label{eq:filteredcomplex}
C_k^{\epsilon_0}  \xhookrightarrow{f^0}  C_k^{\epsilon_1}  \xhookrightarrow{f^1}\cdots \hookrightarrow  C_k^{\epsilon_{i-1}} 		\xhookrightarrow{f^{i-1}}	 C_k^{\epsilon_{i}}  	\xhookrightarrow{f^i} C_k^{\epsilon_{i+1}}	\hookrightarrow \cdots \hookrightarrow	 C_k^{\epsilon_{P}}  
\end{equation}
Each complex is associated with a chain group $ C^{\epsilon_i}_k $. The Rips complex at a finer scale is included in that at a grosser scale, and this inclusion induces a corresponding inclusion map between adjacent chain groups $ f^i\colon C_k^{\epsilon_{i}}\to C_k^{\epsilon_{i+1}} $ as in Eq~\ref{eq:filteredcomplex}. Importantly, these inclusion maps associate the holes in the complexes across scales. Each independent $k$-dimensional hole can then be represented as an interval $ (\epsilon_b, \epsilon_d) $, where $ \epsilon_b $ is the scale at which a hole emerges, i.e.~its \textit{birth scale}, and $ \epsilon_d $ is the scale at which it is filled in, i.e.~its \textit{death scale}. The life span $ \epsilon_d-\epsilon_b $ indicates how persistent the hole is across scales (not to be confused with persistence over time, e.g.~metastable dwells). With this interval representation, we can visualize these $ k^{th} $ homological features across scales as a \textit{barcode} \cite{Ghrist2007}. Fig~\ref{fig:pd}D and E show the persistence of generators of the $ 0^{th} $ and $ 1^{st} $ homology groups $ H_0 $ and $ H_1 $ respectively, capturing connected components and loops across scales. The set of all intervals, the barcode, constitutes a multiscale topological portrait of the point cloud $ X $. Using the software Perseus developed by Nanda \cite{perseus}, we compute two multiscale topological portraits, the $ 0^{th} $ and $ 1^{st} $ persistent homology, for reasons stated in Section \nameref{section:method_topmeta}, for each segment $ X(t) $ of a frequency-phase graph, as described in Section~\nameref{section:method_decomp}, for $ t=2,3,\cdots,48 $. 

\subsection*{Topological recurrence plot}\label{section:method_toprecur}
To study the dynamics of metastable patterns, we need to construct the recurrence plot of multiscale topological portraits. This requires us to define a measure of distance between any two such portraits. In the present study, we use \textit{persistence landscape distance} as a metric, considering its low computation time and potential for statistical use \cite{Bubenik2017}. Persistence landscapes \cite{Bubenik2015} essentially convert a barcode (Fig~\ref{fig:pd} D, E) to a sequence of piecewise-linear functions (Fig~\ref{fig:pd} F, G). The construction of persistence landscapes requires two steps: (1) representing the persistence of each topological feature as a tent function that rises from zero to peak during the first half of its life and falls back to zero during the second half, and (2) taking envelopes of those tent functions in a nested manner. More specifically, let's consider a barcode of $B$ bars, i.e. a set of birth-death intervals $\{ (\epsilon_b^{(i)},\epsilon_d^{(i)}) \}_{i=1}^B$. The goal is to convert them into a sequence of piecewise-linear \textit{landscape functions} $ \{\lambda^{(\ell)}\}_{\ell=1}^L $. Intervals are first used to construct a sequence of tent functions
\begin{equation}
\sigma_i(\epsilon)=\begin{cases}
0 & \epsilon \notin (\epsilon_b^{(i)}, \epsilon_d^{(i)})\\
\epsilon-\epsilon_b^{(i)} & \epsilon\in (\epsilon_b^{(i)},\frac{\epsilon_b^{(i)}+\epsilon_d^{(i)}}{2}]\\
\epsilon_d^{(i)} - \epsilon & \epsilon \in (\frac{\epsilon_b^{(i)}+\epsilon_d^{(i)}}{2},\epsilon_d^{(i)})
\end{cases}
\end{equation}
for $ i=1,2,\cdots,B $. Define $ \lambda^{(\ell)}(\epsilon) $ to be the $ \ell^{th} $ largest value of $ \{\sigma_i(\epsilon)\}_{i=1}^B $. The smaller the $ \ell $, the more prominent the features captured by the landscape function $\lambda^{(\ell)}$. For example, Fig~\ref{fig:pd}F, G show the first five landscape functions computed from intervals in D, E respectively (the infinite interval that appears in every barcode is ignored in the computation, see Fig~\ref{fig:pd}D). Note that for each $\epsilon$ the sequence $ \lambda^{(\ell)}(\epsilon)$ is decreasing, and  $ \lambda^{(L)} $ is the smallest function that is not zero for all $ \epsilon $. Now we can compare multiscale topological portraits as functions. We define the distance between the $ k^{th}$-level persistent homology of two point clouds $ X $ and $ X' $ as the supremum norm of the difference between their corresponding average landscape functions,
\begin{equation}
D_k(X,X')=\|\bar{\lambda}_{k}(\epsilon)-\bar{\lambda}'_{k}(\epsilon)\|_\infty=\sup\limits_\epsilon|\bar{\lambda}_{k}(\epsilon)-\bar{\lambda}'_{k}(\epsilon)| \label{eq:D_k}
\end{equation}
where $ \bar{\lambda}_{k}(\epsilon)=\frac{1}{L}\sum_{\ell=1}^L\lambda_{k}^{(\ell)}(\epsilon) $, and $ \{\lambda_k^{(\ell)}\}_{\ell=1}^L $ and $ \{\lambda_k'^{(\ell)}\}_{\ell=1}^L $ are the landscape functions of the $ k^{th}$-level persistence homology of $X$ and $X'$. 

With a metric defined, we are now in a position to compute the recurrence plot of a multiscale topological portrait, a \textit{topological recurrence plot}, which is a distance matrix with components $ d_{i,j}=D_k(X(t_i),X(t_j)) $ for the $ k^{th} $ persistent homology of segments $ X(t) $ of a frequency-phase graph (Fig~\ref{fig:recur_pd}A-B, D-E). The subdiagonal of this matrix reflects the rate of change of multiscale topological portraits as a function of time. 

To provide a direct comparison between topological and non-topological recurrence, we define a \textit{pointwise} metric by treating each segment of a frequency-phase graph, a point-cloud with $ M $ points in 3D, as a state vector with $ 3M $ components, 
\begin{align}
d_x(X(t_i),X(t_j))=\bigg\| &\bigg((x_1-t_i)-(x'_1-t_j),\cdots,(x_M-t_i)-(x'_M-t_j), \nonumber\\
&\dfrac{W\left(2\pi(x_{M+1}-x'_{M+1})\right)}{2\pi},\cdots,\dfrac{W\left(2\pi(x_{2M}-x'_{2M})\right)}{2\pi}, \nonumber\\
&x_{2M+1}-x'_{2M+1},\cdots,x_{3M}-x'_{3M} \bigg)^\intercal \bigg\|\label{eq:d_x}
\end{align}
for any two segments $ X(t_i)=\{x_m\}_{m=1}^M $ and $ X(t_j)=\{x'_m\}_{m=1}^M  $ , where the function $ W $ wraps its argument to the interval $ (-\pi,\pi] $ by $W(z)=\Arg(e^{iz})$ and $ \|\cdot\| $ is the $ L_2 $-norm. This metric treats the movement of each point independently, and an associated recurrence plot may be called \textit{a pointwise recurrence plot}. In contrast to the recurrence plot of relative phase (e.g. Fig~\ref{fig:recur_phi}), the pointwise recurrence plot of segments of a frequency-phase graph is a more appropriate non-topological counterpart of the topological recurrence plot as defined above. The difference between topological and pointwise recurrence plots directly demonstrates the value of multiscale topological analysis that is not attributable to frequency-phase decomposition or inclusion of a temporal neighborhood alone. 

In the next section, we show how topological recurrence plots help reveal transitions in the coordination dynamics between eight agents (Fig~\ref{fig:eight_example}) that eluded traditional methods of visualization and analysis (Fig~\ref{fig:eight_example}A, Fig~\ref{fig:recur_phi}B). 

\section*{Applying topological recurrence to empirical data}
\label{section:results}

\subsection*{Validation with simple dynamics}
\begin{figure}[H]
	\begin{adjustwidth}{-2.25in}{0in}
	\centering
	\includegraphics[width=\linewidth]{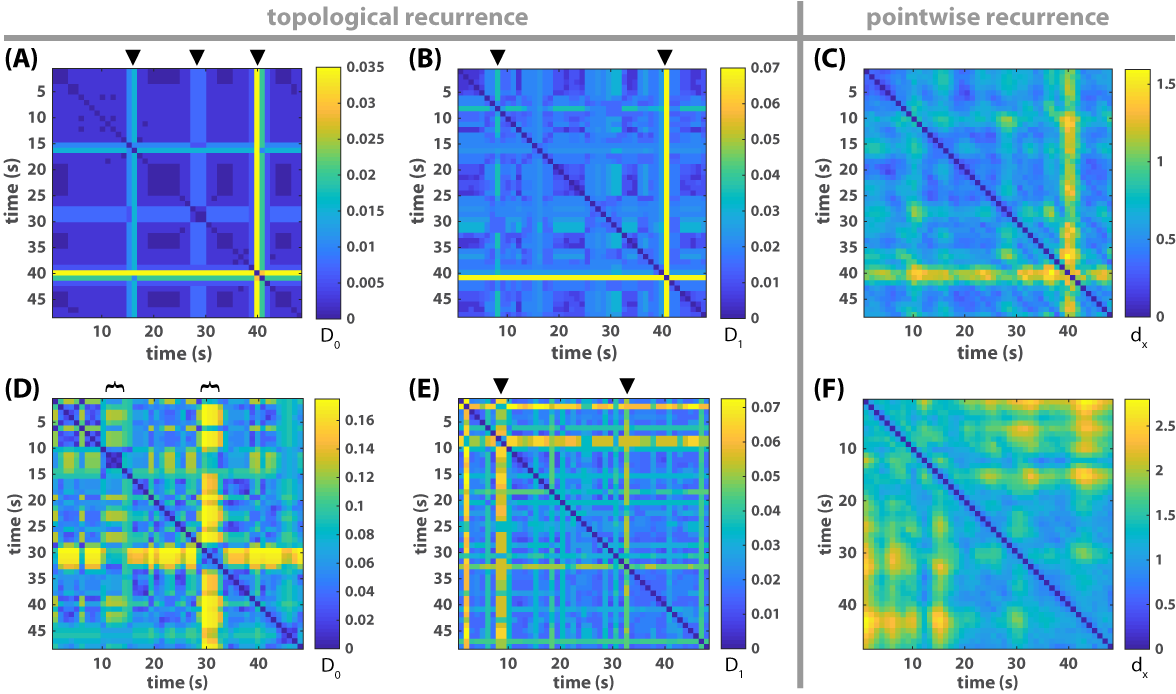}
	\caption[Recurrence plots of topological features versus states.]{
		{\bf Topological versus pointwise recurrence plots of metastable dynamics}. (A-B) shows the recurrence of connected components and loops respectively for the triadic example of Fig~\ref{fig:triad_example}, where the color of each pixel indicates the topological distance between segments of the frequency-phase graph at time $ x $ and time $ y $, as defined in Eq~\ref{eq:D_k}, i.e. $ D_0(X(x),X(y)) $ for (A), $ D_1(X(x),X(y)) $ for (B). Black triangles on top mark the time of topological transitions. They correspond very well with transitions in the original relative phase dynamics (Fig~\ref{fig:triad_example}A) and its associated recurrence plot (Fig~\ref{fig:recur_phi}A). (C) shows the pointwise recurrence plot of the triadic example, where the color of each pixel reflects the distance between point clouds $ X(x) $ and $ X(y) $ as state vectors, as defined in Eq~\ref{eq:d_x}, instead of their multiscale topological portraits. Similar transitions also appear in (C) as in (A-B) though less sharp. (D-F) shows the corresponding recurrence plots for the eight-agent example of Fig~\ref{fig:eight_example}. In the recurrence plot of connected components (D), two transitions are apparent, each of which lasts about 5s (marked by black brackets). The onset of the first transition (around 10s) and the offset of the second (about 33s) also stand out in the recurrence plot of loops (E), marked by black triangles. These features are not apparent in the pointwise recurrence plot of the frequency-phase graph (F), or the recurrence of relative phase (Fig~\ref{fig:recur_phi}B). 
	} \label{fig:recur_pd}
	\end{adjustwidth}
\end{figure}

Before investigating the eight-agent example, we first validate this method with the triadic example (Fig~\ref{fig:triad_example}), the dynamics of which we already know (see Section~\nameref{section:examples}). Fig~\ref{fig:recur_pd}A, B shows the topological recurrence plots of connected components ($ 0^{th} $ persistent homology landscape) and loops ($ 1^{st} $ persistent homology landscape) respectively. In the recurrence plot of connected components (Fig~\ref{fig:recur_pd}A), three moments stand out against the background (marked by black triangles), indicating sudden changes in connected components. These topological changes successfully identify the escapes from three recurring all-inphase patterns in the original relative phase dynamics (steepened magenta trajectory $ \phi_{13} $ between black bars in Fig~\ref{fig:triad_example}A; the same interruptions in the coordination pattern can be seen in Fig~\ref{fig:recur_phi}A). The recurrence plot of loops (Fig~\ref{fig:recur_pd}B) reveals a small transition around 10s and a large transition around 40s (marked by black triangles) in terms of loops. They capture the disturbances surrounding the first formation of the all-inphase pattern and its eventual destruction (see oscillation of frequency trajectories around 10 and 40s in Fig~\ref{fig:triad_example}B). Taken together, topological recurrence (Fig~\ref{fig:recur_pd}A, B) is able to faithfully capture important transitions of coordination patterns observed in the triadic example (Fig~\ref{fig:triad_example}), as much as the traditional recurrence plot of the relative phase (Fig~\ref{fig:recur_phi}A). For comparison, a pointwise recurrence plot of the frequency-phase graph is shown in Fig~\ref{fig:recur_pd}C, which also captures the essential transitions but not as definitively as the topological recurrence plots (Fig~\ref{fig:recur_pd}A, B) or the recurrence plot of relative phases (Fig~\ref{fig:recur_phi}A). This indicates that the clarity present in the topological recurrence plots (Fig~\ref{fig:recur_pd}A, B) is not solely due to the use of the frequency-phase graph (Section~\nameref{section:method_decomp}) instead of relative phases. 

\subsection*{Topological recurrence plots reveal structures in complex coordination patterns}\label{section:toporecur}
Following the basic validation above, we applied the same analysis to the eight-agent example from Fig~\ref{fig:eight_example}, where coordination patterns are more complex and the transitions between them remain obscure under traditional means of analysis. Here, the topological recurrence plots (Fig~\ref{fig:recur_pd}D-E) are strikingly structured, compared to the original dynamics (Fig~\ref{fig:eight_example}), the recurrence plot of relative phases (Fig~\ref{fig:recur_phi}B) or the pointwise recurrence plot of the frequency-phase graph (Fig~\ref{fig:recur_pd}F). The recurrence plot of connected components (Fig~\ref{fig:recur_pd}D) shows a major transition around 30s and a minor one around 10s (marked by black brackets on top of Fig~\ref{fig:recur_pd}D). The onset of the 10s transition and the offset of the 30s transition are also highlighted by the transition of loops (marked by triangles in Fig~\ref{fig:recur_pd}E). Next we return to the original relative phase and frequency dynamics (Fig~\ref{fig:eight_explain}) to investigate what underlies these topological transitions, with an emphasis on the transitions of connected components.

\subsection*{Topological transitions reflect collective change in original dynamics}
\begin{figure}[H]
	\begin{adjustwidth}{-2.25in}{0in}
	\centering
	\includegraphics[width=\linewidth]{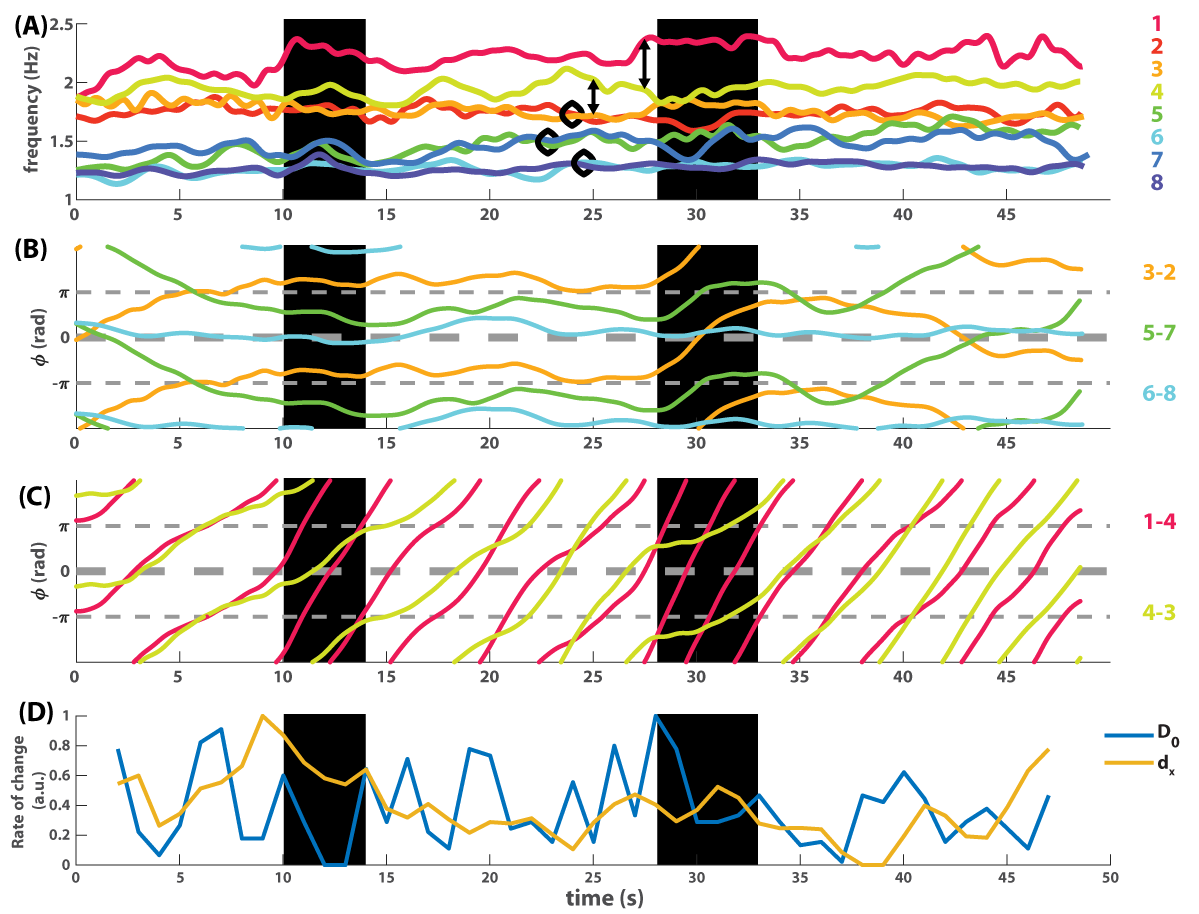}
	\caption[Breaking down the frequency, relative phase and topological dynamics of the eight-agent coordination.]{
		{\bf Examining the dynamic relevance of topological transitions in the eight-agent example}. (A-C) shows trajectories of frequency and relative phase from Fig~\ref{fig:eight_example}, smoothed out by a 2s moving average. (A) shows the frequency dynamics of all eight agents. (B) shows the dynamics of three slowly varying relative phases (thickened trajectories in Fig~\ref{fig:eight_example}A), corresponding to three pairs of frequency trajectories enclosed by black circles in (A). (C) shows the dynamics of two fast varying relative phases (among the thin trajectories in Fig~\ref{fig:eight_example}A), corresponding to relations between frequency trajectories connected by double arrows in (A). (D) shows the rate of change of connected components (blue trajectory) and pointwise rate of change (yellow trajectory), which is the distance between two consecutive multiscale topological portraits under the metric $ D_0 $ (Eq~\ref{eq:D_k}) and $ d_x $ (Eq~\ref{eq:d_x}) respectively. Both trajectories are normalized by mapping $[\min,\max]\mapsto [0,1] $ for comparison. Two transitional periods seen in Fig~\ref{fig:recur_pd}D are highlighted with black backgrounds, bordered by adjacent peaks in the blue trajectory in (D). See text for interpretation. 
	}\label{fig:eight_explain}
	\end{adjustwidth}
\end{figure}

We found that, indeed, transitions in the topological recurrence plot detect meaningful changes in the original dynamic patterns. We focus on the most prominent transition of connected components around 30s (Fig~\ref{fig:recur_pd}D, right bracket). The corresponding original dynamics is shown in Fig~\ref{fig:eight_explain}, highlighted by a black rectangle around 30s. Right before this transition, the ensemble was in a relatively stable configuration with three frequency pairs (Fig~\ref{fig:eight_explain}A, trajectories enclosed by black circles), a lone wolf (agent 1, magenta trajectory on top in Fig~\ref{fig:eight_explain}A), and a commuter oscillating between its neighbors (agent 4, yellow trajectory in Fig~\ref{fig:eight_explain}A). At the onset of the transition (28s), two pairs suddenly broke up (3-2, 5-7), switched partners with others, eventually returning to the original configuration at the offset of the transition (33s). Importantly, this transition is also reflected in the dynamics of the relative phases (Fig~\ref{fig:eight_explain}BC). In particular, the transition took exactly the time for pair 3-2 (orange trajectory in Fig~\ref{fig:eight_explain}B) to be destabilized from antiphase and return to antiphase after wrapping around for one cycle, suggesting a collective transition may be controlled by local dynamics. Importantly, such a collective transition is only revealed by topological recurrence, and not by traditional means of analysis. 

The minor transition of connected components (10s) marks a more local event -- the splitting of the higher frequency group (warm colors Fig~\ref{fig:eight_explain}A) in to a lone wolf (agent 1, magenta in Fig~\ref{fig:eight_explain}A), a pair (3-2, orange and red in A), and a commuter in between (yellow in A). The transitions of loops (marked by triangles in Fig~\ref{fig:recur_pd}E), on the other hand, highlight the local disturbances accompanying these changes in coordination patterns. For example, the transition of loops around 10s (Fig~\ref{fig:recur_pd}E, first triangle) reflects high amplitude oscillation in the frequencies of agents 1 and 3 (magenta and orange trajectories in Fig~\ref{fig:eight_example}B and Fig~\ref{fig:eight_decomp}B) before the sudden departure of agent 1. 

Overall, topological recurrence plots reveal local and global transitions of coordination patterns in the data that elude traditional methods. To have a sense of why non-topological recurrence is insensitive to such information, we compare side by side the topological rate of change (blue trajectory in Fig~\ref{fig:eight_explain}D) and the pointwise rate of change (yellow trajectory). We find that greater topological changes (peaks of blue trajectory) generally do not require a large pointwise change (e.g. yellow trajectory at 28s) , and conversely, large pointwise changes (peaks of yellow trajectory) do not imply a large topological change (e.g, blue trajectory at 9s). In other words, topological recurrence captures some of the interdependency between the movement of individual points in the point cloud that is irreducible to the aggregate of the independently measured movements of each point. Such irreducibility of certain topological properties may tap into the very nature of collective transitions in multiscale coordinative structures.  

\section*{Discussion}\label{section:discuss}

The present work introduces a multiscale topological approach to understanding metastable coordination between many diverse agents. We first gave a conceptual framework of how the study of metastable phase coordination can be converted into a multiscale topological analysis of frequency graphs. Under this conceptual framework, we employed persistent homology as a natural analytical tool for such multiscale topological problems. We further demonstrated for proof of concept how topological recurrence plots helped uncover structures and transitions in example trials of multiagent social coordination from \cite{Zhang2018}, especially those that were elusive to traditional means (e.g. contrast topological recurrence plots in Fig~\ref{fig:recur_pd}D, E with its traditional counterpart in Fig~\ref{fig:recur_phi}B). In particular, an important topological transition (major transition around 30s in Fig~\ref{fig:recur_pd}D, \ref{fig:eight_explain}) was discovered, showing how sudden, coordinated pattern switching can occur across multiple local groups segregated in frequency. Taken together, the conceptual and computational tools developed here provide a new perspective on the analyses of complex rhythmic patterns with multiscale and metastable characteristics, an advance that has been deemed to be much necessary \cite{Tognoli2014Enlarging,Tognoli2014}.

In addition to its utility in characterizing metastable rhythmic coordination, a few important features of this topological approach may prove advantageous in more general settings regarding coordination phenomena in complex systems. First, a simplicial complex obtained as a Rips complex associated to a point cloud is a purely relational, coordinate-free representation of the structure of the point cloud, and hence is invariant under rigid motions as is its homology. In the study of coordination phenomena, relational quantities are essential \cite{Kelso1995, Kelso2012}. These relational quantities and their evolution in a dynamic pattern may be conveniently described 
through the topology of simplicial complexes. Second, topology is the mathematical discipline that integrates local information into global information \cite{Thom1975structural}. In multiagent coordination, the number of relational quantities multiplies, for example there are $ N(N-1)/2 $ dyadic relations between $N$ agents. These relational quantities constrain each other and form higher-order structures which may not be discernible by examining each quantity individually. Algebraic topology provides a mature set of tools to build global pictures from such local (e.g.~dyadic) relational quantities. In the present work, the ability of topological portraits to capture global properties is a key to detecting transitions in collective patterns that are not just an accumulation of pointwise changes (see Fig~\ref{fig:eight_explain}D and corresponding text). In other words, topological portraits capture emergent features in the collective dynamics that are not reducible to the sum of its parts, tapping into a key feature of complex systems \cite{Holland1998,Wolfram2002,miller2009complex,Kelso2009}. Last but not least, persistent homology \cite{Edelsbrunner2002,Zomorodian2005} provides a well-developed mathematical framework to describe multiscale structures. The coexistence of multiple relevant scales happens to be a central characteristic of complex systems (from the biochemical to the social), making meaningful methods of multiscale analyses highly valuable (e.g.~\cite{Simon1977,Oltvai2002,Bar-Yam2004,Sales-Pardo2007,Vespignani2012,Sekara2016,Aguilera2018}). In short, the topological approach outlined in the present work may serve as a prototype for more general analyses of complex systems.  

Over the past decade, computational topology has gradually attracted the attention of biologists as a set of new tools to shed light on geometrical or topological structures in complex, high-dimensional data that were difficult to quantify or visualize by traditional means. For example, various types of topological portraits (not limited to persistent homology) have been used to study the shape of viral evolutionary tracks \cite{Chan2013}, RNA folding pathways \cite{Yao2009}, collective encoding of global spatial organization by groups of neurons \cite{Curto2008,Dabaghian2012} and the geometry of neural dynamics \cite{Petri2014,Saggar2018} (see \cite{Giusti2016} for more applications in neuroscience). In contrast to these studies, where topological portraits were the primary subject of analyses and interpretation, here we focused on the \textit{change} of topological portraits without direct interpretation of the portraits per se. Given the time of topological changes (e.g. brackets in Fig~\ref{fig:recur_pd}D), we returned to the original dynamics to verify that those topological changes truly reflect transitions between different phase coordination patterns. There are both technical and theoretical reasons to proceed in this fashion. Technically, the direct interpretation of a topological portrait may be affected by how the original data were sampled and preprocessed (e.g. Section~\nameref{section:method_decomp}). But it is much less ambiguous to interpret changes in the topological portrait: If we see a change in the topological portrait, there must be a corresponding change in the coordination pattern; and if the change in the coordination pattern is sufficiently small, the change in the topological portrait is also small, due to the fact that the topological portraits are stable with respect to the metric defined in Eq~\ref{eq:D_k} \cite{Bubenik2015}. On the theoretical side, studying transitions is a key to understanding nonlinear dynamical processes; and for reasons shown in Section~\nameref{section:method_topmeta}, one would expect changes in the multiscale topological portrait during transitions within and between metastable patterns. Furthermore, topological features during a transition per se may be highly nongeneric compared to what happens before or after a transition, and therefore stand out in the topological recurrence plot (e.g. Fig~\ref{fig:recur_pd}D). For these reasons, the current method directly addresses dynamically relevant topological changes in the original time series, without interpreting the topological portraits as an intermediate step. At this stage of development of a multiscale topological approach to complex collective patterns, it helps to check whether one's system of measurement really captures the dynamic features of interest. 

The present approach, however, does not exclude a direct analysis and interpretation of the topological portraits. Indeed, the motivation behind using persistent landscapes is that they are conducive to further statistical analyses. Before that we need to know more concretely which quantities are relevant to the kind of dynamic features of interest. For example, does the magnitude of change in connected components reflect in general how global a transition is? Can we accurately classify different types of transitions by taking into account more statistical features of the landscape functions? To deliver valid answers, potential methods may be validated based on transitions whose classification we know a priori, e.g. simulated transitions based on mathematical models of biological coordination (e.g.~\cite{Zhang2019jrsi} for a model developed based on the Human Firefly experiment \cite{Zhang2018}). 

To conclude, we presented a multiscale topological approach to understanding the metastable coordination dynamics involving multiple agents. We demonstrated by analysis of examples of dynamics and theoretical discussions that this method possesses great potential for characterizing complex, multiscale dynamic patterns. Further developments using simulated time series are desirable to move toward a systematic method of classifying phase transitions in complex collective dynamics, for which this work provides a prototype.

\section*{Acknowledgments}
The authors of this paper were supported by NIMH (MH-080838) and NIBIB (EB-025819).
\nolinenumbers

\bibliography{FireflyTDA_arXiv_v09182019} 

\section*{Supporting information}
\beginsupplement

\paragraph{Simulations of metastable coordination}\label{SItext:simulation}
Examples of metastable dynamics shown in Fig~\ref{fig:explain_freqtopo} are simulated following the equation
\begin{equation}
	\dot{\varphi}_i=\omega_i-a\sum_{j=1}^{N}\sin (\varphi_i-\varphi_j)-b\sum_{j=1}^{N} \sin 2(\varphi_i-\varphi_j) \label{eqn:NHKB}
\end{equation}
where $ \varphi_i $, $ \dot{\varphi}_i $, and $ \omega_i $ are the phase, the time derivative of phase, and the natural frequency of the $ i^{th} $ oscillator in an ensemble of $ N $, respectively, and $ a,b\geq 0 $ are the first-order and second-order coupling strengths between oscillators. This system \cite{Zhang2019jrsi} has been shown to capture key experimental observations of coordination between multiple people as observed in \cite{Zhang2018}. For the example shown in Fig~\ref{fig:explain_freqtopo}~A-B, $ \omega_{1,2}=0, 0.6 $ Hz and $ a=b=1 $ with zero initial phases. For the example shown in Fig~\ref{fig:explain_freqtopo}~C-D, the average natural frequency $ \bar{\omega}=1.5 $ Hz, the natural frequency difference between two adjacent oscillators $ \omega_{i+1}-\omega_{i}=0.075 $ Hz, and overall coupling strength $ a=b=0.15 $ with random initial conditions.

\end{document}